# ANALYZING THE IMPACT OF VISITORS ON PAGE VIEWS WITH GOOGLE ANALYTICS

MOHAMMAD AMIN OMIDVAR[1], VAHID REZA MIRABI[2] AND NARJES SHOKRY[3]

[1]Faculty of Information Technology Management, Iran Islamic Azad University E-campus
Amin.omidvar@gmail.com

[2]Faculty of Management, Islamic Azad University Central branch (Tehran), Iran
vrmirabi@yahoo.com

[3]Faculty of Public Administration, Islamic Azad University Central branch (Tehran), Iran
nargesshokry@yahoo.com

## ABSTRACT

*This paper develops a flexible methodology to analyze the effectiveness of different variables on various dependent variables which all are times series and especially shows how to use a time series regression on one of the most important and primary index (page views per visit) on Google analytic and in conjunction it shows how to use the most suitable data to gain a more accurate result.*

*Search engine visitors have a variety of impact on page views which cannot be described by single regression. On one hand referral visitors are well-fitted on linear regression with low impact. On the other hand, direct visitors made a huge impact on page views. The higher connection speed does not simply imply higher impact on page views and the content of web page and the territory of visitors can help connection speed to describe user behavior. Returning visitors have some similarities with direct visitors.*

## 1. Introduction

The internet is growingly rapidly and has a great impact on many businesses. Thousands of companies now own a website and websites have become an integrated part of the business. Furthermore many companies have employed many technologies which are available through the web such as online services. With web information, web developers and designers can improve user interfaces, search engines, navigation features, online help and information architecture and have happier visitors/ costumers [21]. One of the most popular ways which most frequented websites use to collect data and information about their websites is through web analytic. Web analytic collects a large amount of data from users such as browser type, connection speed, screen size, visitors' type, etc. The collected data are usually large in quantity and type that need to be further processed to become useful information or knowledge.

## 2. Literature review

The internet has been playing the important role of corporate marketing during the past ten years [43]. With its combination of rich text, multimedia and user involvement, the internet contains more information than any other media [26], [38]. The internet offers speed, reach, and multimedia advantages, and has changed the way in which businesses interact with their customers, suppliers, competitors, and employees [12].

Nearly all businesses now have a website [17]. A corporate website enriches the image of a business and provides direct benefits in terms of electronic commerce (e-commerce) sales [30] and indirect benefits in terms of information retrieval, branding, and services [34]. Recognition of the internet is driving marketers in traditional companies to conduct transactions on the internet [15]. Barua, Konana, B.Whinston, & yin (2001) found that e-business operational excellence results in financial performance [8]. Thousands of companies have a fear to be left behind by their competitors if they do not use online technologies.

The total number of internet users and the number of websites are increasing significantly (table 1-1, table 1-2) which will result in the rapid growth of the use of the world wide web for commercial purposes.

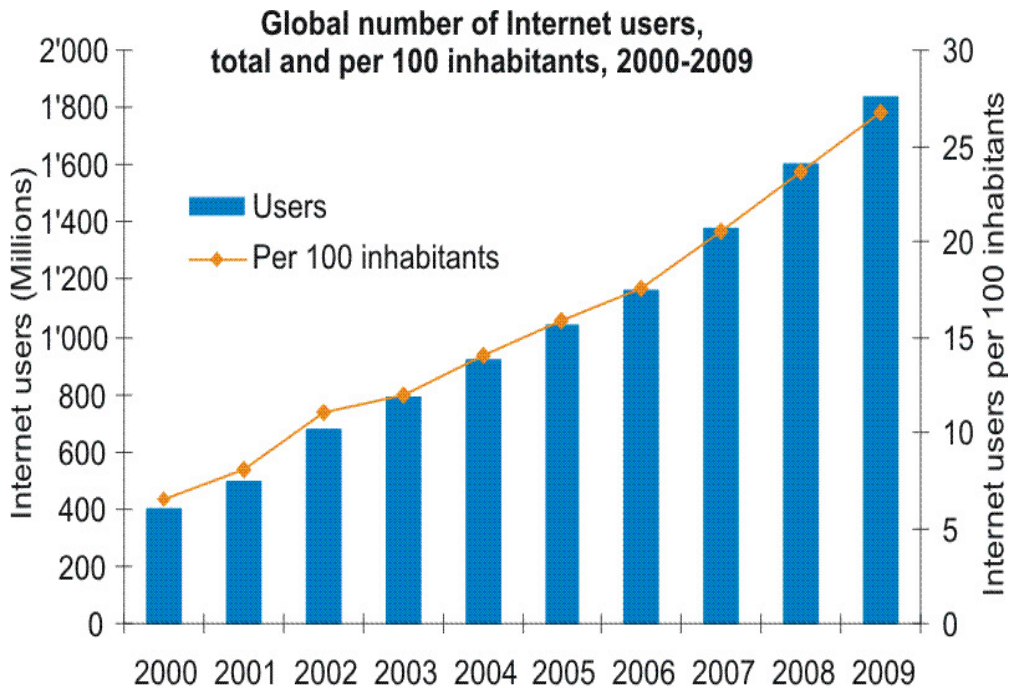

Figure 1, (Global Number of Internet Users,total and per 100 inhabitants 2000-2009)

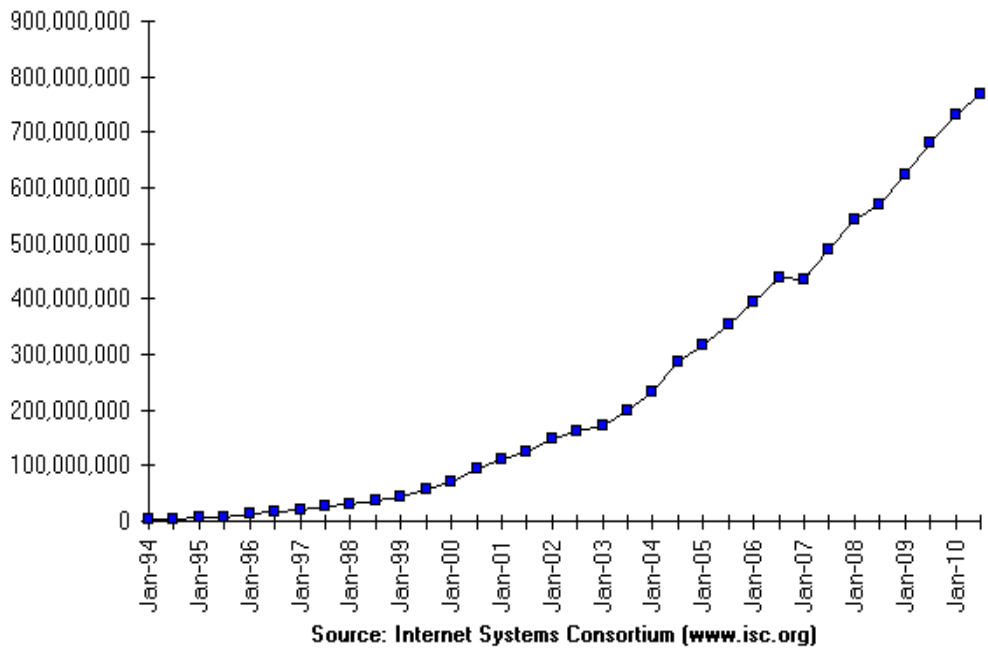

Figure 2 Internet Domain Survey Host Count(Consotium, 2010)

On a five year forecast by Forrester, e-commerce sales in the U.S. will grow 10% annually from 2014 and online retail sales will be nearly $250 billion, up from $155 billion in 2009 [40].

The increasing amount of internet users, websites and retail sales implies that web developing should be carried out in a competent, professional manner to increase profit. However, systematic analysis of costumers' behaviors has not kept pace with the rapid growth in e-commerce. Without quantifiable metrics which is available through web analytics software, website optimization (WSO) is a guessing game, therefore a majority of e-commerce companies cannot afford this risk given their huge amount of money. Above 70% of the most frequented websites use web analytic tools but with their large amount of data, it is difficult to use them effectively[1]. Therefore, it is important to understand what kind of data and knowledge are required for successful website development work.

Web Analytics Association Standards (2006) committee defined the three most important metrics as Unique Visitors, Visits/Sessions, and Page Views; and, also categorized search engine marketing metrics through counts (visits...), ratios (page views per visits....), and key performance indicators (KPIs) [5].

The main reasons for measuring Search engine marketing (SEM) successes are related to traffic measurement and the return on investment come on 4th [41]. The top for reasons are as follows:

- Increased traffic volume (76%)
- Conversion rates (76%)
- Click-through rates or CTRs (70%)
- Return on investment (67%)

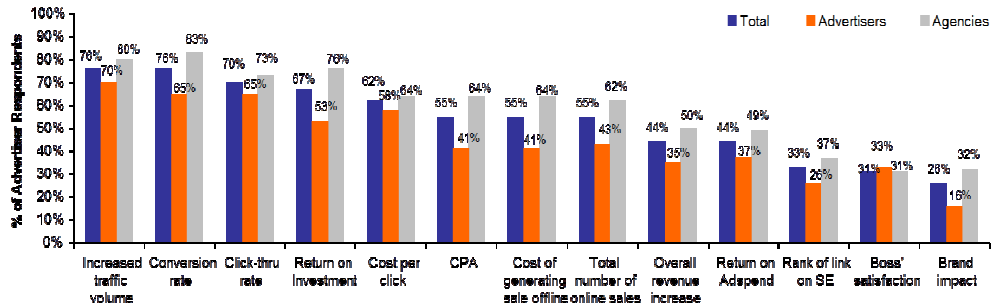

Figure 3 Metrics used to measure Search Engine Marketing campaign success

Progressive improvement of SEM campaigns, conversion rates, and website performance are available through web metrics, which would results in an increase in profits, happier customers, and higher return on investment (ROI) by tracking progress over time or against the competition [11].

Online technology collects large amounts of detailed data on visitor traffic and activities on websites, which would cause a plethora of metrics [20], and on the other hand this variety of measures can be overwhelming. Developing a website is a dynamic ongoing process which is guided by knowledge of its visitors.

## 2. Profile of the website

In 1998 an Iranian visual artist website was launched (http://www.omidvar.net). This website has many pages with images and few texts. The Google Analytics traffic overview showed that all traffic sources sent a total of 19,703 visits from 1 June 2008 to 1 May 2010. The total page views during this period were 100,144.

## 3. Methodology

Google Analytics allows users to export report data in Microsoft Excel format, which when transformed can be analyzed with time series statistical programs. The software EViews is used to compute time series regression. Initially a data set with 19,703 entries for 23 months drawn from Google Analytics was employed to analyze the performance of page views or page views per visits. Monthly data series was the most suitable series among daily and weekly because the accuracy and credibility of the regression was higher than those of other series.

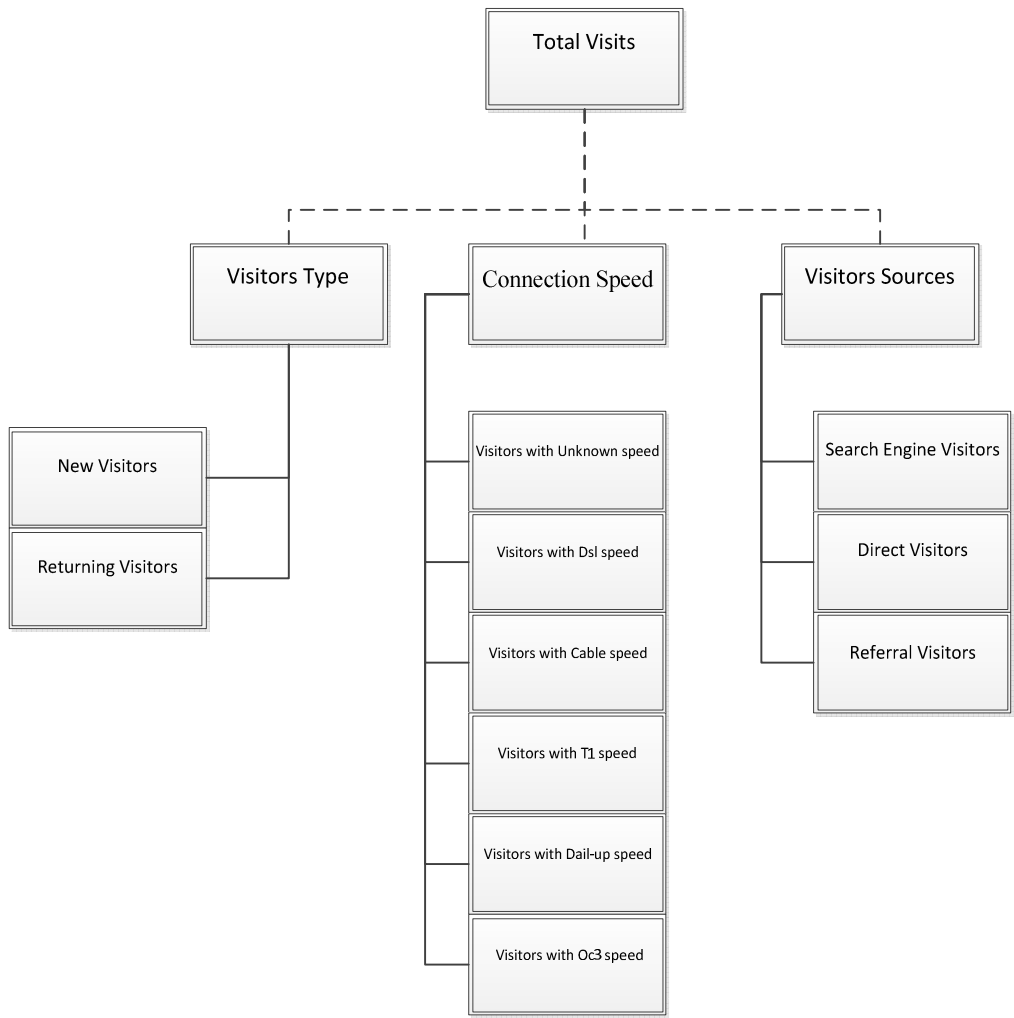

Figure 4 Independent Variables

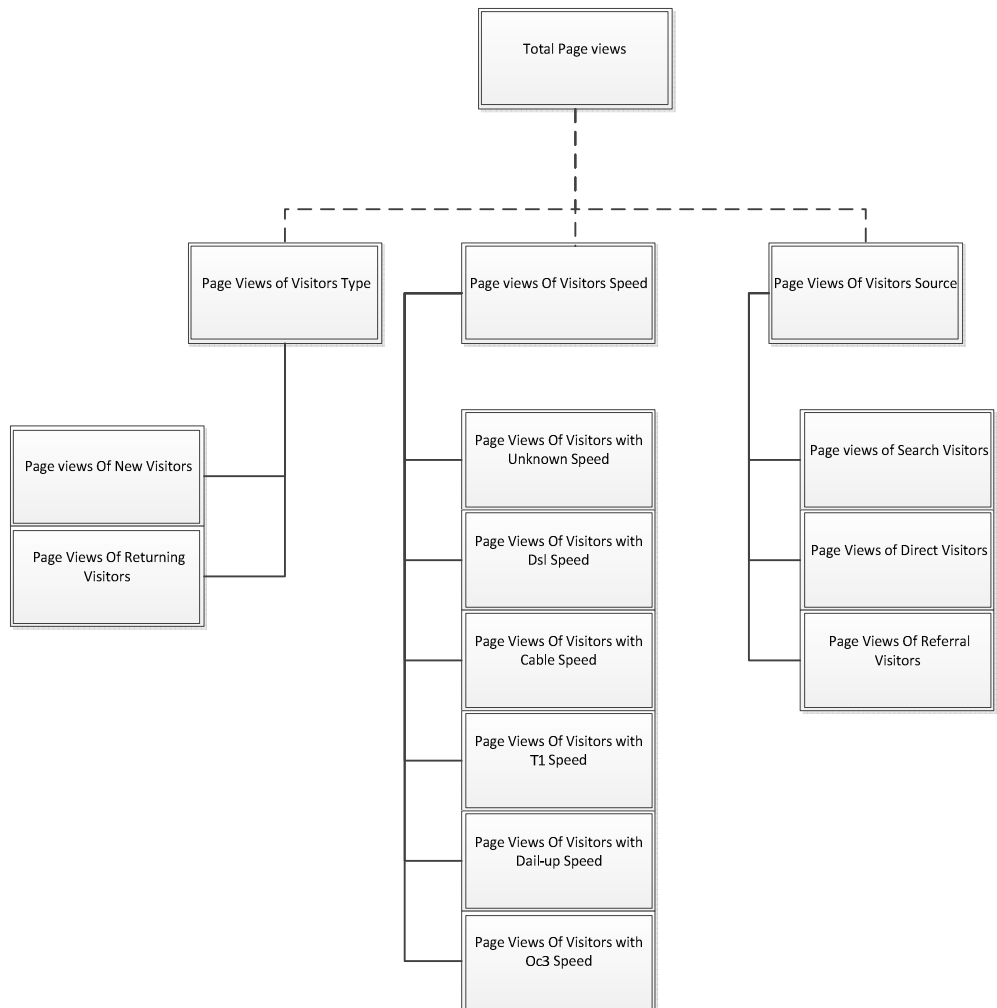

Figure 5 Dependent Variables

Before the hypothesis testing, some statistical matters with regards to the use of Google Analytics data in combination with time series methodology must be considered. For this reason all independent variables were processed by Augmented Dickey-Fuller Test to see if they were stationary or not. If the variables had a unit root they would be transformed to stationary by Difference-Stationary Process.

Firstly, the total page views were broken down to its elements. For example page views of new visitors and returning visitors are elements of total page views regarding to page views of visitors type.

Secondly, Autoregressive Moving Average (ARMA) model were created with their independent variables and their dependent variables.

Thirdly, a new ARMA model of total page views was created with the sum of its related fundamental model. With the combined use of both dependent variables and independent variables, more accurate results were achieved.

Finally, the new model credibility and reliability was checked which is consistent with these seven steps [27].

1) Regression line must be fitted to data strongly ($R^2>0.5$).
2) Independent variables should be jointly significant to influence or explain the dependent variable (i.e. F-test, Anova).
3) Most of the independent variables should be individually significant to explain dependent variable (i.e. T-test).
4) The sign of the coefficients should follow economic theory or expectation or experiences or intuition.
5) No serial or auto-correlation in the residual (Breusch-Godfrey serial correlation LM test : BG test)
6) The variance of the residual (u) should be constant meaning that homoscedasticity (Breusch-Pegan-Godfrey Test)
7) The residual (u) should be normally distributed (Jarque Bera statistics).

# 4. Hypothesis testing

The main questions are:

- Are page views well described by visitors' sources? Which source has the highest impact and why?
- Are page views well described by visitors' connection speed? Which connection speed has the highest impact and why?
- Are page views well described by visitors' type? Which type has the highest impact and why?
- This survey starts with a preparation stage which checks if our independent variables are stationary series or not and transforms them to stationary.

| Independent Variable | Stationary Series | Difference-Stationary Process |
|---|---|---|
| Search Visitors | Yes | |
| Direct Visitors | No | Yes |
| Referral Visitors | Yes | |
| New Visitors | Yes | |
| Returning Visitors | Yes | |
| Visitors with Unknown Speed | Yes | |
| Visitors with DSL Speed | Yes | |
| Visitors with Cable Speed | Yes | |
| Visitors with T1 Speed | Yes | |
| Visitors with Dial-Up Speed | Yes | |
| Visitors with OC3 Speed | Yes | |

Table 1 Augmented Dickey Fuller (ADF) Test and Difference-Stationary Process

## 4.1 The impacts of visitors' sources on page views:

As Table 1 illustrates, Direct Visitors are not stationary series therefore they are transformed to stationary series by difference-stationary process and the new variable is called "Direct Visitors*".

In the Introduction it was mentioned that there are a variety of data in web analytics and the metric values should be chosen wisely, therefore the independent variable (page view) is broken down to the page views of direct visitors, referral visitors and search visitors as shown in Table2.

| Independent Variable | Dependent variable | R^2 | Probe value | Coefficient |
|---|---|---|---|---|
| Page views of Search Visitors | Search Visitors | 0.41 | 0 | 3.82 |
| Page views of direct Visitors | Direct Visitors* | 0.3 | 0.01 | 4.44 |
| Page views of Referral Visitors | Referral Visitors | 0.61 | 0 | 1.77 |

Table 2 ARMA Regression for Page Views Sources

Direct Visitors* was transformed with Difference-Stationary Process to stationary series, therefore the amount of R^2 which is lower than 0.5 is not an issue to show the fitness of model to data; but on the other hand, for search visitors which its R^2 is lower than 0.5 implies that the regression is unlikely to be a good estimate.

Page views = page views of search engine visits + page views of direct visits + page views of referral visits

Page views = 3.82*Search Visitors + 4.24*Direct Visitors + 1.77*Referral Visitors + c

| Independent Variable | Dependent variables | Regression Fitted to Data Strongly | Jointly Significant | Coefficients Follow Economic Theories | No Serial in the Residual | Homoscedasticity of the Residual Variance | Residual Normality Distributed |
|---|---|---|---|---|---|---|---|
| Total page views | Visitors sources | Yes | Yes | Yes | No | Yes | Yes |

Table 3 Assumptions of good Regression model of total page views with visitors sources

Excepting the serial in the residual this regression is a good model and shows the impact of Direct and Referral visitors correctly.

## 4.2 The impacts of visitors' connection speed on page views:

The dependent variables of visitors connection speed are broken down and an ARMA module for each is created as follows.

| Independent Variable | Dependent variable | R^2 | Probe value | Coefficient |
|---|---|---|---|---|
| Page Views of Visitors with Unknown Speed | Visitors with Unknown Speed | 0.69 | 0 | 4.81 |
| Page Views of Visitors with DSL Speed | Visitors with DSL Speed | 0.391 | 0 | 2.74 |

---

[1] Unlikely to be a good estimate

| Page Views of Visitors with Cable Speed | Visitors with Cable Speed | 0.74 | 0 | 1.71 |
|---|---|---|---|---|
| Page Views of Visitors with T1 Speed | Visitors with T1 Speed | 0.47 | 0 | 2.22 |
| Page Views of Visitors with Dial-Up Speed | Visitors with Dial-Up Speed | 0.41 | 0 | 3.16 |
| Page Views of Visitors with OC3 Speed | Visitors with OC3 Speed | 0.53 | 0 | 7.81 |

Table 4 ARMA Regression for Page Views Connection Speed

Visitors with DSL, T1 and Dial-Up speed are unlikely to be a good estimate because their $R^2$ is lower than 0.5.

Page views = page views of Unknown visitors + page views of DSL Visitors + page views of Cable Visitors+ page views of Dial-Up Visitors+ page views of OC3 Visitors

Page Views=4.81*Unknown visitors +2.74*DSL Visitors +1.71* Cable Visitors+2.22*T1+3.16*Dial-Up Visitors+7.81*OC3 Visitors +C

| Independent Variable | Dependent variables | Regression Fitted to Data Strongly | Jointly Significant | Coefficients Follow Economic Theories | No Serial in the Residual | Homoscedasticity of the Residual Variance | Residual Normality Distributed |
|---|---|---|---|---|---|---|---|
| Total page views | Visitors Connection Speed | yes | yes | yes | yes | yes | yes |

Table 5 Assumptions of good Regression model of total page views with visitors Connection Speed

## 4.3 The impact of visitors' type on page views:

| Independent Variable | Dependent variable | R^2 | Probe value | Coefficient |
|---|---|---|---|---|
| Page Views of New Visitors | New Visitors | 0.68 | 0 | 2.40 |
| Page Views of Returning Visitors | Returning Visitors | 0.70 | 0 | 5.22 |

Table 6 ARMA Regression for Page Views Type

Page Views= Page Views of New visitors+ Page Views of Returning Visitors
Page Views= 2.40*New visitors+ 5.22*Returning Visitors

| Independent Variable | Dependent variables | Regression Fitted to Data Strongly | Jointly Significant | Coefficients Follow Economic Theories | No Serial in the Residual | Homoscedasticity of the Residual Variance | Residual Normality Distributed |
|---|---|---|---|---|---|---|---|
| Total page views | Visitors Connection Speed | yes | yes | yes | yes | yes | yes |

Table 7 Assumptions of good Regression model of total page views with visitors Type

## 5. Results and Analysis

The results of this survey are categorized in two categories:
1. Results from Regression
2. The methodology that is used

### 5.1 Results from Regression:

Most of the direct visitors have visited the site before and memorized the address or entered the site from traditional channels so namely they could be fan of the website, the website owners and the website developers or visitors from traditional channels (conference, meeting, and event). Those visitors who access the website from traditional channels would probably increase during the time of event or meeting and those visitors who are site owners or developers will probably increase visits during the update or develop time. Consequently these visitors were not stationary and had unit root.

**Analyzing the impact of visitors sources on pages views**

By each direct visitor, the number of page views was increased to 4.44 and for referral visitors it was increased to 1.77; therefore the direct visitors had a higher impact in comparison to referral visitors. Additionally, exchanging links with more related websites do improve referral visitors' effect on page views. The Page views of search visitors were not fitted to the module and varied very much. In order to understand search visitors behavior, their quarries meaning should be understood which is available through semantic web technologies. Investigation of keywords' priority and bases on discarded certain percentage of web pages hence making the search more compact and efficient [25].

**Analyzing the impact of visitors connection speed on pages views**

By literature review it was thought that the higher speed connection would cause a greater impact on page views, but things are not so simple. This is due to access by Dial-Up visitors which used a lower connection speed and had more impact on page views than cable and T1 visitors. It is believed that the content type on the region of visitors plays an important role on that. Since this site belongs to an Iranian artist, it must have a higher impact on visitors from Iran.

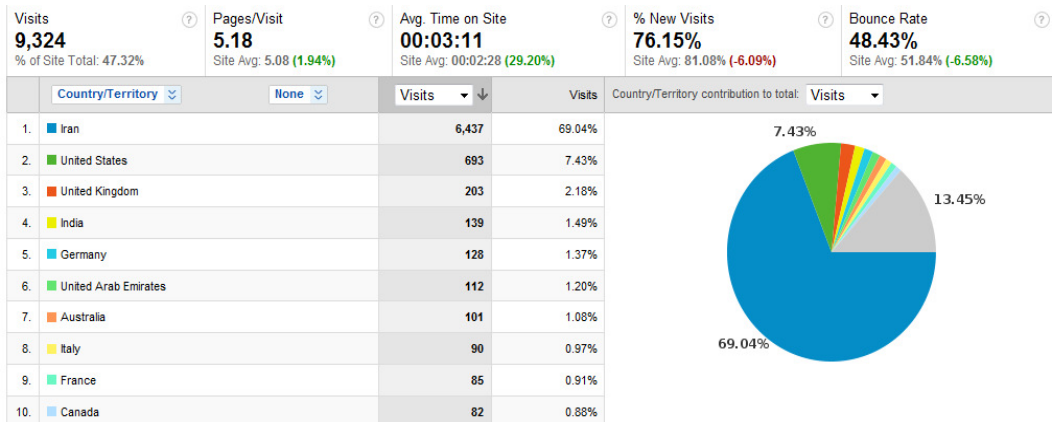

Figure 6 Percentage of Territory of Unknown visitors

Figure 6 reveal that Unknown visitors are mainly from Iran therefore it proves their high impact on page views.

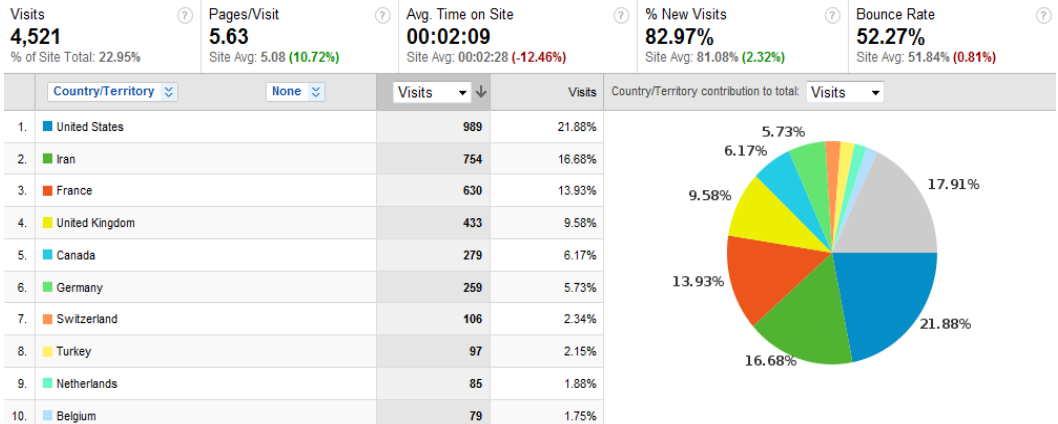

Figure 7 percentage of Territory of DSL Visitors

17% of DSL visitors which on average had a speed of 1.5 Mbps were from Iran. That is the reason why it had more impact on page views than Cable visitors with average speed of 4 Mbps.

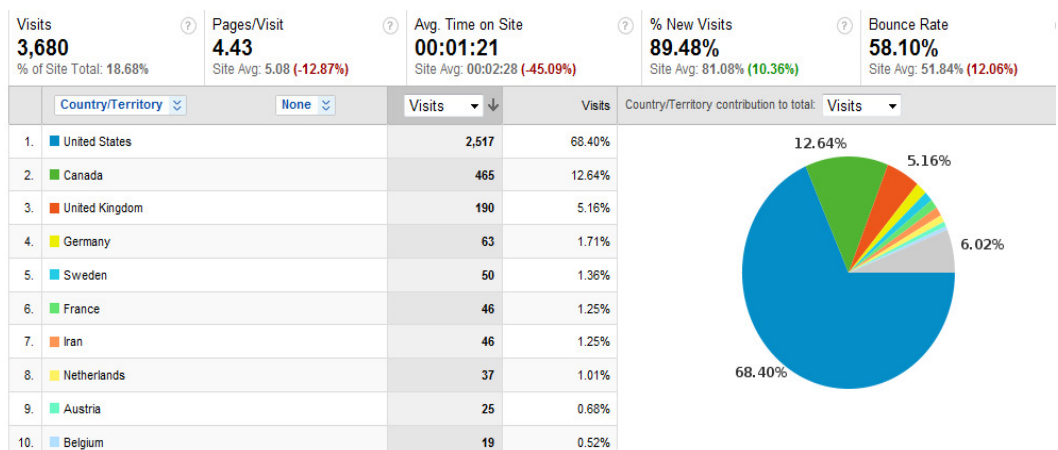

Figure 8 Percentage of territory of Cable visitors

Cable visitors had the lowest impact although their average speed was around 4 Mbps. This can be explained by the low percentage of Iranian visitors.

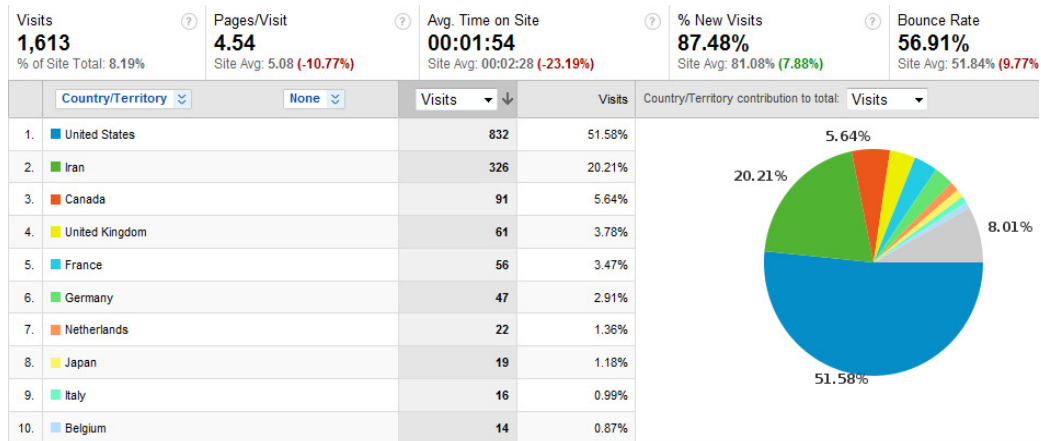

Figure 9 Percentage of Territory of T1 visitors

20.21% of T1 visitors are from Iran and the connection speed roughly would be 1.5 MBPS. It is expected that these visitors had the same impact as DSL visitors; and, indeed their impact was close to each other (2.22 for T1 visitors and 2.74 for DSL visitors).

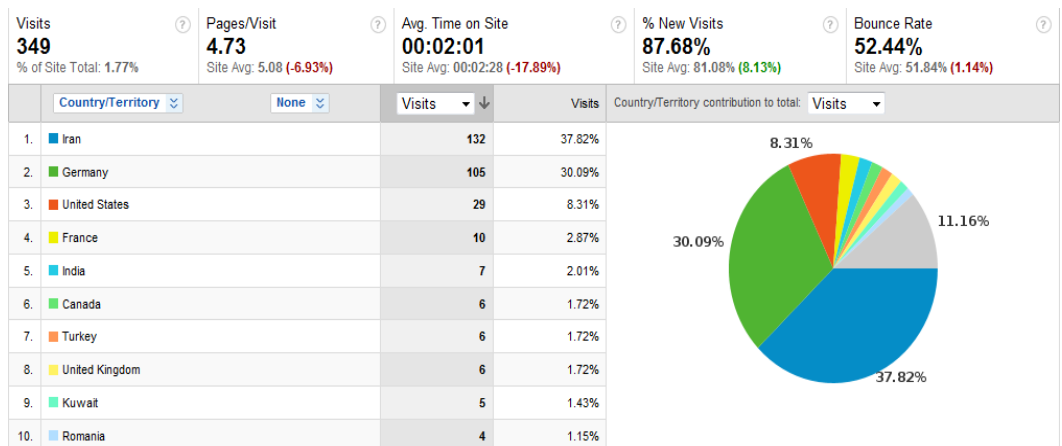

Figure 10 Percentage of Territory of Dial-Up visitors

Dial-Up visitors are mainly from Iran so it does prove their impact (3.16) on page views. The Oc3 visitors had the highest speed which on average is 155 MBPS; therefore their impact on page views was as high as 7.81.

In conclusion, what can be manifest from close analysis of these data is that it is easier to predict user behavior whose territory and connection speed does not varied much.

**Analyzing the impact of visitors Type on pages views**

Returning visitors had more impact on page views than new visitors did and it was quite high. Returning visitors must have similarities with direct visitors because all of the direct visitors excepting visitors from traditional channels had visited the site before. So their impact must be close to each other (5.22 for returning visitors and 4.24 for direct visitors).

### 5.2 The New Methodology

Many websites have measured their success with their competitors' behavior and many others have measured that with their own website success. This survey had introduced a methodology to measure the success of the website with its time series data. It also focused on one of the most primary and important variables which are page views and showed how to use the most suitable data for that. This method can be used on all websites and time series variables.

## Recommendation

It is suggested to use this methodology on some other artists' sites, and other variables such as a combination of these variables. Furthermore since Google Analytics is integrated with Adwords and Adsense and their time series data's are available on Google Analytics, further studies on time series data of Adsense Revenue or Adwords campaigns might have interesting results; because they bring visitors and revenue to the site.

This methodology is not recommended for search visitors' behavior and needs further investigation. In order to understand search visitors behavior their quarries meaning should be understood which is available through semantic web technologies. Investigation of keywords' priority and classification would make the search more compact and efficient [25].

According to user referral link, territory and the meaning of the content, the ontology can be simulated by Protégé which is a free, open source platform for constructing, visualizing and manipulating domain models and knowledge based applications with ontologies [35].

## Discussion

Many website owners or developers have used web traffic elements for their website performance, but the important thing is the knowledge gained about visitors and their behavior to keep them happy and satisfied with the website. Most of the websites have used web analytics to collect data about visitors with no systematic way to convert these data into tangible knowledge. The amount of these data which is available through web analytic is immense, and the developer may get lost in it. So a systematic way is needed to analyze these data.

Firstly, this survey used time series regression with dependent variables and independent variables. And all these variables were chosen wisely to have the most accurate result.

Secondly, the total page views module was created with sum of detailed regression module. For example for new and returning visitors, the total page views were created as follow:
Page views of New visitors= New visitors*2.40 + C
Page views of Returning visitors= Returning visitors*5.22 + C
Total page views= Page views of new visitors + Page views of Returning visitors + C

Finally, seven assumptions as listed by Hossain were checked as the final module to see how good the ARMA module was [28].

In sum, this paper showed how to use a regression module with the most suitable dependent and independent variables and its functionality.

## APPENDICES

### Augmented Dickey-Fuller Unit Root Tests:

The variables under the study are the following: Direct visits (DV), reference site visits (RSV), search engine visits (SV), new visits (NV), returning visits (RV), visitors with unknown speed (UNKNOWNV), visitors with DSL Speed (DSLV), visitors with Cable Speed (CABLEV),visitors with T1 Speed (T1V), visitors with Dial-Up Speed (DIAL-UPV) and visitors with OC3 Speed (OC3V).

DV ADF Test Statistic:-2.85 10 percent Critical Value:-2.65
RSV ADF Test Statistic:-2.15 10 percent Critical Value:-2.64
SV ADF Test Statistic:-2.85 10 percent Critical Value:-2.65
SV* ADF Test Statistic:-4.48 10 percent Critical Value:-2.65
NV ADF Test Statistic:-3.95 10 percent Critical Value:-2.64
RV ADF Test Statistic:-2.74 10 percent Critical Value:-2.64
UNKNOWNV  ADF Test Statistic:-2.73 10 percent Critical Value:-2.64
DSLV  ADF Test Statistic:-3.62 10 percent Critical Value:-2.64
CABLEV  ADF Test Statistic:-4.51 10 percent Critical Value:-2.64
T1V  ADF Test Statistic:-3.65 10 percent Critical Value:-2.64
DIAL-UPV  ADF Test Statistic:-3.49 10 percent Critical Value:-2.64
OC3V  ADF Test Statistic:-2.2 10 percent Critical Value:-1.60

## ARMA Regression:

Dependent Variable: PAGEVIEWS
Method: Least Squares
Date: 10/06/10   Time: 12:28
Sample (adjusted): 2008M07 2010M04
Included observations: 22 after adjustments
Page views=3.82*Search Visitors+4.24*Direct
Visitors+1.77*Referral Visitors+ c

|  | Coefficient | Std. Error | t-Statistic | Prob. |
|---|---|---|---|---|
| C(1) | 2609.016 | 256.2849 | 10.18014 | 0.0000 |
| R-squared | 0.581917 | Mean dependent var | | 4252.682 |
| Adjusted R-squared | 0.581917 | S.D. dependent var | | 1859.102 |
| S.E. of regression | 1202.083 | Akaike info criterion | | 17.06589 |
| Sum squared resid | 30345061 | Schwarz criterion | | 17.11548 |
| Log likelihood | -186.7248 | Hannan-Quinn criter. | | 17.07757 |
| Durbin-Watson stat | 0.947281 | | | |

Table 8 ARMA Regression for Sources

Dependent Variable: PAGEVIEWS
Method: Least Squares
Date: 10/07/10   Time: 20:02
Sample (adjusted): 2008M06 2010M04
Included observations: 23 after adjustments
Page Views=4.81*Unknown visitors +2.74*DSL
Visitors +1.71* Cable
Visitors+2.22*T1+3.16*Dial-Up
Visitors+7.81*OC3 Visitors +C

|  | Coefficient | Std. Error | t-Statistic | Prob. |
|---|---|---|---|---|
| C(1) | 1332.591 | 198.0689 | 6.727918 | 0.0000 |

| R-squared | 0.742444 | Mean dependent var | 4346.913 |
|---|---|---|---|
| Adjusted R-squared | 0.742444 | S.D. dependent var | 1871.734 |
| S.E. of regression | 949.9051 | Akaike info criterion | 16.59311 |
| Sum squared resid | 19851032 | Schwarz criterion | 16.64248 |
| Log likelihood | -189.8207 | Hannan-Quinn criter. | 16.60552 |
| Durbin-Watson stat | 1.560109 | | |

Table 9 ARMA Regression for connection speed

Dependent Variable: PAGEVIEWS
Method: Least Squares
Date: 06/28/10   Time: 21:54
Sample (adjusted): 2008M06 2010M04
Included observations: 23 after adjustments
PAGEVIEWS=2.4*A__+5.22*B__+C(1)

|  | Coefficient | Std. Error | t-Statistic | Prob. |
|---|---|---|---|---|
| C(1) | 1843.713 | 212.7060 | 8.667892 | 0.0000 |

| R-squared | 0.702971 | Mean dependent var | 4346.913 |
|---|---|---|---|
| Adjusted R-squared | 0.702971 | S.D. dependent var | 1871.734 |
| S.E. of regression | 1020.102 | Akaike info criterion | 16.73570 |
| Sum squared resid | 22893393 | Schwarz criterion | 16.78507 |
| Log likelihood | -191.4605 | Hannan-Quinn criter. | 16.74811 |
| Durbin-Watson stat | 1.549155 | | |

Table 10 ARMA Regression for Types

## Breusch-Godfrey Serial Correlation LM Test (BG Test)

Breusch-Godfrey Serial Correlation LM Test:

| F-statistic | 3.558785 | Prob. F(2,19) | 0.0487 |
|---|---|---|---|
| Obs*R-squared | 5.995448 | Prob. Chi-Square(2) | 0.0499 |

Table 11LM Test on ARMA model of Sources

Breusch-Godfrey Serial Correlation LM Test:

| | | | |
|---|---|---|---|
| F-statistic | 1.115946 | Prob. F(2,20) | 0.3472 |
| Obs*R-squared | 2.309003 | Prob. Chi-Square(2) | 0.3152 |

Table 12 LM Test on ARMA model of Connection speed

Breusch-Godfrey Serial Correlation LM Test:

| | | | |
|---|---|---|---|
| F-statistic | 1.293293 | Prob. F(2,20) | 0.2963 |
| Obs*R-squared | 2.633929 | Prob. Chi-Square(2) | 0.2679 |

Table 13LM Test on ARMA model of Type

## Breusch-Pegan-Godfrey Test:

Heteroskedasticity Test: Breusch-Pagan-Godfrey

| | | | |
|---|---|---|---|
| F-statistic | 0.065414 | Prob. F(3,18) | 0.9775 |
| Obs*R-squared | 0.237265 | Prob. Chi-Square(3) | 0.9714 |
| Scaled explained SS | 0.288654 | Prob. Chi-Square(3) | 0.9621 |

Table 14Breusch-Pagan-Godfrey test on model of sources

Heteroskedasticity Test: Breusch-Pagan-Godfrey

| | | | |
|---|---|---|---|
| F-statistic | 2.066530 | Prob. F(6,16) | 0.1153 |
| Obs*R-squared | 10.04188 | Prob. Chi-Square(6) | 0.1229 |
| Scaled explained SS | 9.298063 | Prob. Chi-Square(6) | 0.1575 |

Table 15Breusch-Pagan-Godfrey test on model of Connection Speed

Heteroskedasticity Test: Breusch-Pagan-Godfrey

| | | | |
|---|---|---|---|
| F-statistic | 1.060743 | Prob. F(2,20) | 0.3649 |
| Obs*R-squared | 2.205738 | Prob. Chi-Square(2) | 0.3319 |
| Scaled explained SS | 1.809690 | Prob. Chi-Square(2) | 0.4046 |

Table 16Breusch-Pagan-Godfrey test on model of Types

## Normality Test for the residuals

Jarque-Bera Test for visitors' Source: 3.92 Prob: 0.14
Jarque-Bera Test for visitors' Connection Speed: 3.29 Prob: 0.19
Jarque-Bera Test for visitors' Type: 2.48 Prob: 0.29

**Authors**

Mohammad Amin Omidvar received his master information technology management degree from Islamic Azad University (E-campus), Tehran, Iran, in 2010. He has had many researches in E-commerce field and as an innovator he has deployed information technology in various fields especially web, multimedia and digital painting which has been presented in many Iranian universities' speeches and conferences.

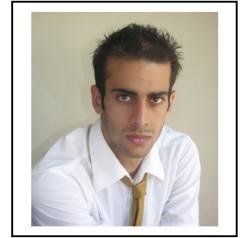

Vahid Reza Mirabi has PH.D in marketing management and is working as a vice president of management school and was the president of Islamic Azad university Tehran central branch. He has more than 16 years of experience in academic and industry in Iran. He received his PH.D from GHARWAL Srinegar from Poona University in India. He is advisor of many companies in the field of marketing. He has published 5 books and several articles. His major area of interest is marketing research, strategic management and human resource management.

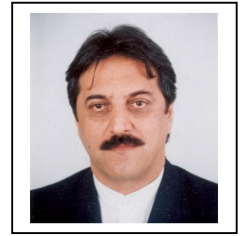

Narjes Shokri is working as a head department of public administration (management) and has more than 10 years of experience in academics and more than 15 years of experience in teaching. She received BA and MA in public administration at Tehran University and PH.D at Islamic Azad University Tehran. Her major area of interest is method of research, MIS and organizational behaviors. She has authored several researches in national journal.

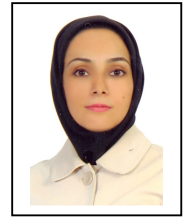